\documentclass[conference]{IEEEtran}
\IEEEoverridecommandlockouts
\usepackage{cite}
\usepackage{amsmath,amssymb,amsfonts}
\usepackage{algorithmic}
\usepackage{graphicx}
\usepackage{textcomp}
\usepackage{fancyhdr}
\usepackage{comment}
\usepackage{xcolor}
\def\BibTeX{{\rm B\kern-.05em{\sc i\kern-.025em b}\kern-.08em
    T\kern-.1667em\lower.7ex\hbox{E}\kern-.125emX}}
\begin{document}

\title{\textbf{SEAM}: A \textbf{SE}cure \textbf{A}utomated and \textbf{M}aintainable Smart Contract Upgrade Framework}

\author{\IEEEauthorblockN{Tahrim Hossain$^{1}$, Faisal Haque Bappy$^{2}$, Tarannum Shaila Zaman$^{3}$, and Tariqul Islam$^{4}$}
\IEEEauthorblockA{
$^{1, 2, 4}$ Syracuse University and
$ ^{3}$ University of Maryland, Baltimore County (UMBC)\\
Email: mhossa22@syr.edu, fbappy@syr.edu, zamant@umbc.edu} and mtislam@syr.edu
} 


\maketitle

\thispagestyle{fancy}
\lhead{This work has been accepted at the 2025 IEEE Consumer Communications \& Networking Conference (CCNC 2025)}
\cfoot{}

\begin{abstract}
This work addresses the critical challenges of upgrading smart contracts, which are vital for trust in automated transactions but difficult to modify once deployed. To address this issue, we propose SEAM, a novel framework that automates the conversion of standard Solidity contracts into upgradable versions using the diamond pattern. SEAM simplifies the upgrade process and addresses two key vulnerabilities: function selector clashes and storage slot collisions. Additionally, the framework provides tools for efficiently deploying, modifying, and managing smart contract lifecycles. By enhancing contract security and reducing the learning curve for developers, SEAM lays a robust foundation for more flexible and maintainable blockchain applications.


\end{abstract}


\section{Introduction}
Many transactions that require trust rely on smart contracts, which are automated agreements executed without needing intermediaries \cite{szabo1996smart}. However, once these contracts are deployed, they cannot be easily changed. Although there are some methods \cite{Proxy-hunting} to make them upgradable, these solutions are often inefficient. One major issue is that developers face a steep learning curve, particularly when dealing with complex concepts like storage layout and pointers in Solidity \cite{Maintenance}. Additionally, many developers use open-source code from platforms like Etherscan and GitHub, which can contain vulnerabilities that are hard to fix without replicating flaws \cite{Kondo}. Existing tools focus on detecting security issues such as re-entrancy attacks \cite{Pan}, storage slot collisions \cite{Ruaro2024NotYT}, and gas inefficiencies \cite{Ghaleb}. However, most don’t fully address the challenges of contract upgrades. In this work, we identify and address two key vulnerabilities in smart contract upgrades and propose SEAM, a framework designed to enable secure, automated, and maintainable contract upgrades. SEAM simplifies the process of upgrading smart contracts by utilizing the diamond pattern \cite{eip_2535}, ensuring flexibility and scalability without risking the integrity of the contract. The following are the major contributions of this work, 
\begin{itemize}
    \item We propose SEAM, a novel framework that automates the conversion of standard Solidity contracts into upgradable ones using the diamond pattern.
    \item SEAM integrates mechanisms to detect and mitigate common vulnerabilities in upgradable smart contracts.
    \item SEAM offers tools for efficiently deploying, modifying, and managing the entire lifecycle of smart contract upgrades.
\end{itemize}

\section{Vulnerabilities in Smart Contract Upgrade}
Upgradable smart contracts provide essential flexibility but come with critical vulnerabilities that developers must recognize to prevent significant financial losses. One of the most concerning issues is the \textbf{function selector clash}, which occurs when multiple functions across different contracts share the same 4-byte identifier derived from their names and parameters. This clash can lead to incorrect function execution, especially in upgradable contracts using \textit{DELEGATECALL} patterns, resulting in unexpected behaviors and security risks as new functions are added during upgrades.
\begin{figure}[h]
   \centering
   \includegraphics[width=\columnwidth]{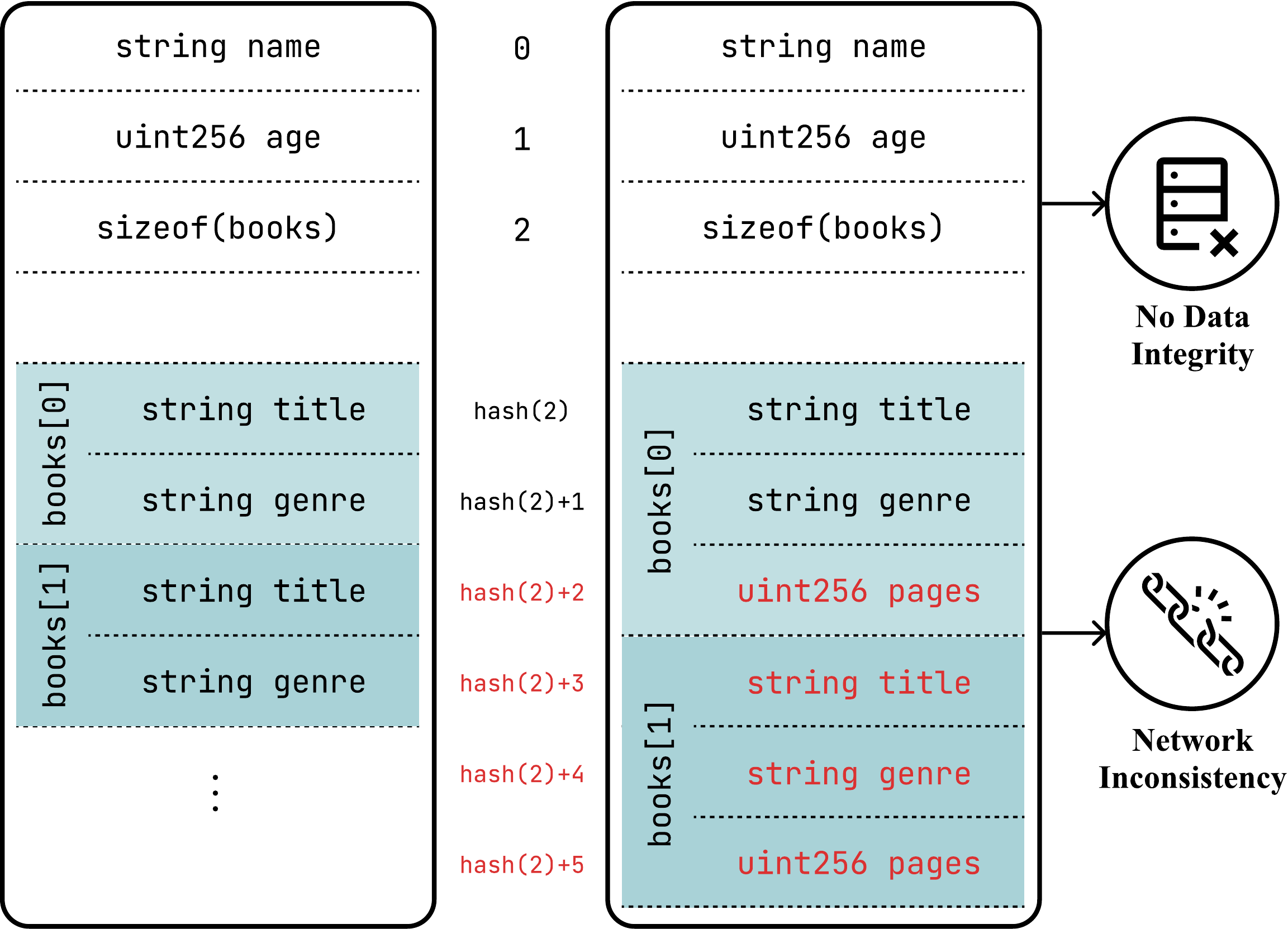}
   \caption{Storage Layout Comparison Before and After Upgrade: Highlighting Potential Slot Collisions in Upgradable Smart Contracts.} 
   \label{fig:slot_collision}
\end{figure}
Another significant vulnerability is \textbf{storage slot collision}, which happens when different contracts share the same storage space but interpret the data inconsistently. This issue often arises in upgrade patterns utilizing \textit{DELEGATECALL}, where both the proxy and implementation contracts access shared storage. In figure \ref{fig:slot_collision}, instances of the Book struct are stored one after the other in an array, taking up consecutive storage slots. When a new parameter is added to the Book struct (pages), it disrupts the existing storage layout, leading to a collision. This collision can lead to data corruption, which may compromise data integrity, impair operational reliability, and undermine user trust due to functional failures.

\begin{figure*} [t]
    \centering
    \includegraphics[width=\textwidth]{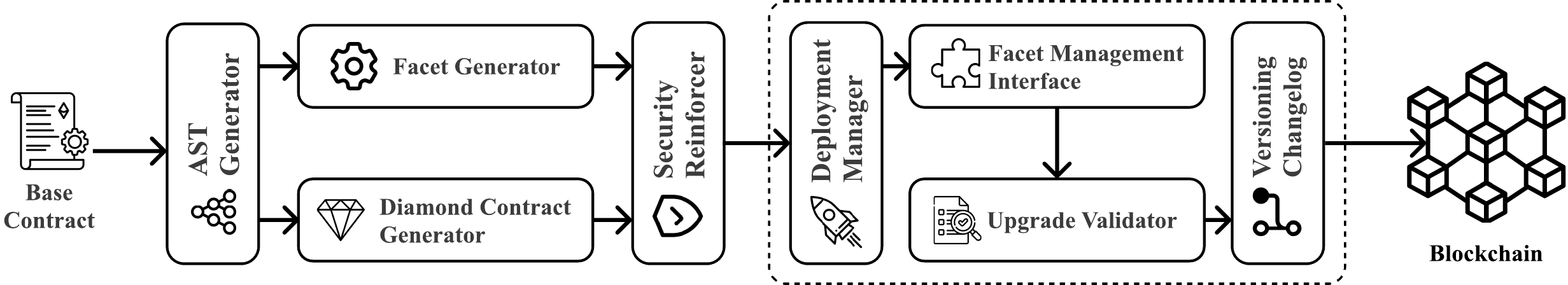}
    \caption{Smart Contract Upgrade Process in SEAM Framework} 
    \label{fig:workflow} 
\end{figure*}

\section{System Architecture of SEAM}
To mitigate the vulnerabilities and potential data corruption, we propose a novel way to upgrade smart contracts securely in an automated and maintainable way. We named our framework SEAM. Figure \ref{fig:workflow} shows the high-level workflow of the smart contract upgrade process in SEAM. 

The workflow begins with the \textbf{AST Generator}, which takes a standard Solidity smart contract as input and generates its Abstract Syntax Tree (AST). This transformation allows for a more structured analysis and modification of the contract's logic. This AST is then passed to the Contract Assembly Unit. Within this unit, the \textbf{Facet Generator} analyzes the AST to modularize the contract's logic, identifying and grouping related state variables and functions based on their dependencies. It arranges the storage layout using the diamond storage pattern to avoid conflicts and extracts shared internal functions into a centralized Solidity library for code reuse. The \textbf{Diamond Contract Generator} creates the main diamond proxy contract

Following this, the \textbf{Security Reinforcer} conducts a thorough security analysis of the generated facets and the diamond contract to identify vulnerabilities like unction selector clashes and storage slot collisions. For function selector clashes, the module traverses the AST to gather all function signatures, ensuring unique mappings by checking for duplicates or conflicting selectors. When conflicts are detected, the module suggests modifications to resolve them. To address storage slot collisions, it identifies scenarios that may lead to conflicts during upgrades—such as arrays of structs and nested structs—and modifies the code accordingly, potentially transforming arrays into mappings or restructuring the layout of struct members. This systematic approach results in a set of modular contracts fortified against common upgrade-related vulnerabilities.

Finally, the \textbf{Deployment Manager} takes care of the last steps in the process. The Deployment Manager sets up the main proxy contract and adds the modular parts, called facets. It also keeps track of these facets and their functions. After everything is set up, the \textbf{Facet Management Interface} allows users to add, replace, or remove facets as needed. Before any upgrades, an \textbf{Upgrade Validator} checks for possible problems like storage slot conflicts and function selector clashes by comparing the new facets with the existing ones. Throughout the contract's life, a \textbf{Versioning and Changelog} system keeps a record of all changes, making it easy to track updates and revert if necessary.

\section{Conclusion}
In this work, we aim to simplify the development and management of upgradable smart contracts while enhancing their security though Our proposed framework, SEAM. We identify and address two key vulnerabilities: function selector clashes and storage slot collisions. In the future, we plan to implement SEAM on a larger scale and evaluate its performance in real-world blockchain environments. By mitigating vulnerabilities and providing tools for efficient upgrade management throughout a contract's lifecycle, we believe SEAM will significantly enhance contract security and streamline upgrades, creating a strong foundation for more flexible and maintainable blockchain applications.


\bibliographystyle{IEEEtran}
\bibliography{IEEEabrv,references}
\end{document}